\documentclass[sigconf,screen]{acmart}

\usepackage{microtype}
\usepackage{xspace}
\usepackage{algorithm}
\usepackage{algpseudocode}
\usepackage{float}
\usepackage{enumitem}
\setlist{leftmargin=*,nosep}
\usepackage{tikz}
\usetikzlibrary{positioning,arrows.meta,fit,backgrounds,shapes.geometric,calc}
\usepackage{listings}
\lstdefinelanguage{json}{
    basicstyle=\ttfamily\small,
    columns=fullflexible,
    showstringspaces=false,
    literate=
     *{0}{{{\color{purple}0}}}{1}
      {1}{{{\color{purple}1}}}{1}
      {2}{{{\color{purple}2}}}{1}
      {3}{{{\color{purple}3}}}{1}
      {4}{{{\color{purple}4}}}{1}
      {5}{{{\color{purple}5}}}{1}
      {6}{{{\color{purple}6}}}{1}
      {7}{{{\color{purple}7}}}{1}
      {8}{{{\color{purple}8}}}{1}
      {9}{{{\color{purple}9}}}{1}
}

\definecolor{colorStrategy}{RGB}{42,114,165}     
\definecolor{colorExecution}{RGB}{191,95,0}      
\definecolor{colorVerification}{RGB}{45,134,45}  
\newcommand{\strategy}{\textcolor{colorStrategy}{\textbf{Strategy}}\xspace}
\newcommand{\execution}{\textcolor{colorExecution}{\textbf{Execution}}\xspace}
\newcommand{\verification}{\textcolor{colorVerification}{\textbf{Verification}}\xspace}
\newcommand{\athena}{\textcolor{colorStrategy}{Athena}\xspace}
\newcommand{\ares}{\textcolor{colorExecution}{Ares}\xspace}
\newcommand{\apollo}{\textcolor{colorVerification}{Apollo}\xspace}

\newcommand{\tbc}{\textsc{TheBotCompany}\xspace}

\newcommand{\finding}[2]{%
  \smallskip\noindent\textbf{Finding~(#1):} \textit{#2}\smallskip}

\newcounter{notesec}[section]

\setcopyright{cc}
\setcctype{by}
\acmDOI{10.1145/3832783.3834358}
\acmYear{2026}
\copyrightyear{2026}
\acmISBN{979-8-4007-2882-2/2026/10}
\acmConference[ASE '26]{Proceedings of the 41st IEEE/ACM International Conference on Automated Software Engineering}{October 12--16, 2026}{Munich, Germany}
\acmBooktitle{Proceedings of the 41st IEEE/ACM International Conference on Automated Software Engineering (ASE '26), October 12--16, 2026, Munich, Germany}
\acmSubmissionID{ase26main-p597-p}
\received{2026-03-26}
\received[accepted]{2026-06-18}

\begin{document}

\title{TheBotCompany: Self-Organizing Multi-agent Systems for Continuous Software Development}

\author{Wenhan Lyu}
\orcid{0009-0004-9129-8689}
\affiliation{%
  \institution{William \& Mary}
  \city{Williamsburg}
  \country{USA}
}
\email{wlyu@wm.edu}

\author{Yue Xiao}
\orcid{0009-0005-7945-464X}
\affiliation{%
  \institution{William \& Mary}
  \city{Williamsburg}
  \country{USA}
}
\email{yxiao05@wm.edu}

\author{Yixuan Zhang}
\orcid{0000-0002-7412-4669}
\affiliation{%
  \institution{William \& Mary}
  \city{Williamsburg}
  \country{USA}
}
\email{yzhang104@wm.edu}

\author{Yifan Sun}
\orcid{0000-0003-3532-6521}
\affiliation{%
  \institution{William \& Mary}
  \city{Williamsburg}
  \country{USA}
}
\email{ysun25@wm.edu}

\begin{abstract}
Large language model (LLM)-based multi-agent systems have shown promise in automating software development tasks. However, most vibe-coding systems focus on completing small tasks and incremental code changes, leaving persistent, continuous software development largely unexplored. We present \tbc\footnote{\href{https://github.com/syifan/thebotcompany}{https://github.com/syifan/thebotcompany}}, an open-source orchestration framework for continuous multi-agent software development. \tbc introduces three key innovations: (1) a three-phase state machine (\strategy $\to$ \execution $\to$ \verification) for milestone-driven development, (2) self-organizing agent teams where manager agents dynamically hire, assign, and retire worker agents based on project needs, and (3) asynchronous human oversight. We evaluate \tbc on real-world software projects over multiple days of continuous development, measuring team adaptation patterns, milestone completion rates, cost efficiency, and code quality. Our results demonstrate that the self-organizing approach enables effective long-term software development with measurable progress, while the verification phase catches defects that would otherwise persist. 
\end{abstract}



\maketitle

\section{Introduction}
\label{sec:intro}

Software development is fundamentally a continuous process~\cite{1456074, 10.1145/2543581.2543595, godfrey2014evolution}. Real-world projects evolve over weeks, months, and years through iterative cycles of planning, implementation, testing, and revision~\cite{59, 1204375, manifesto2001manifesto}. Engineers join and leave teams, priorities shift as requirements change, and the scope of work adapts in response to emerging constraints and discoveries~\cite{10.1145/1355238.1355245, 1182970, 1290455}. Recent Large language model (LLM)-based tools have made rapid progress on code generation and repair~\cite{yang2024swe, bouzenia2025repairagent}, and multi-agent frameworks such as ChatDev~\cite{qian2024chatdev} and MetaGPT~\cite{hong2024metagpt} have shown that organizing LLM agents into role-specialized teams can yield functional software from natural language specifications. However, these systems mostly operate in isolated, session-scoped interactions where a user provides a prompt, the system generates code in minutes, and the session ends with limited ability to maintain context across sessions, track cumulative project-level progress, or resume and iterate on prior work~\cite{10.1145/3643757, 10.1145/3695988, 10449667}. 

We identify three specific gaps in existing approaches. 
\begin{itemize}[leftmargin=*,nosep]
    \item First, \textbf{limited persistent orchestration}: current systems (e.g., ChatDev~\cite{qian2024chatdev}, MetaGPT~\cite{hong2024metagpt}, and single-agent tools such as SWE-Agent~\cite{yang2024swe} and OpenHands~\cite{wang2024openhands}) are primarily designed for single-session execution (each run addresses one task and terminates). While agents can be re-invoked on the same repository, existing frameworks lack built-in mechanisms to track cumulative progress toward milestones, carry forward verification feedback, or automatically resume work after interruptions~\cite{pan2025why, chang2025sagallm, 10.1145/3712003}. The missing coordination state makes long-horizon projects that require iterative refinement over days or weeks difficult to sustain, as the coordination context must be manually reconstructed between sessions. 
    \item Second, \textbf{static or session-scoped team composition}: In ChatDev, MetaGPT, AgileCoder~\cite{nguyen2024agilecoder}, and ALMAS~\cite{tawosi2025almas}, agent roles are predetermined by the framework designer and remain fixed throughout execution. Recent work on dynamic composition---AgentVerse~\cite{chen2024agentverse}, DyLAN~\cite{liu2024dylan}, and Self-Organized Agents~\cite{ishibashi2024self}---adjusts team membership within a single session, but does not persist team adaptations across development cycles or adapt workforce composition in response to verification feedback from prior milestones. Yet real software teams routinely evolve their composition as projects progress---adding a security specialist when a vulnerability surfaces, or retiring a prototyping role once architecture stabilizes~\cite {10.1145/1355238.1355245}. Bridging this gap requires a team composition that is both dynamic \emph{and} persistent across the lifecycle of a project. 
    \item Third, \textbf{synchronous human oversight}: Existing systems support human involvement primarily through synchronous interaction: users either specify requirements before execution, monitor and intervene in real time during a session, or review outputs afterward~\cite{wang2024openhands}. However, real-world software oversight is largely asynchronous, meaning that clients file issues, and managers adjust priorities in a tracker, and developers consume this guidance at natural transition points without blocking ongoing work~\cite{10.1145/3210459.3210471, 10.1145/3712003}. Current frameworks lack a channel for humans to steer multi-day development asynchronously and allows agents to consume guidance at appropriate phase boundaries.
\end{itemize}

To address these gaps, we present \tbc, an open-source orchestration framework for continuous, multi-agent software development. To enable \emph{persistent orchestration}, \tbc organizes development around \emph{milestones}, discrete units of work that progress through a repeating three-phase lifecycle: a \strategy manager defines the next milestone with concrete objectives, an \execution manager assembles and adapts a team of worker agents to implement it, and a \verification manager independently evaluates the result, rejecting incomplete work with structured feedback to trigger corrective cycles. The three-phase design mirrors the plan--implement--review cycle established in iterative software engineering practice~\cite{3766078.3766304} and separates concerns so that no single agent both produces and judges its own output. This lifecycle repeats continuously, with full project state persisted across cycles, driving sustained progress across days of unattended operation. 

To replace \emph{static team composition}, agent teams in \tbc are self-organizing: only the three manager roles are permanent, while managers dynamically hire, specialize, and retire worker agents based on milestone demands. Workers' skills are defined by persistent skill files that managers generate or modify as project needs evolve, enabling the workforce to adapt without framework-level reconfiguration. 

To support \emph{asynchronous human oversight}, \tbc provides a web-based monitoring platform through which human users can inspect project progress, review agent reports and communications, and file issues that manager agents consume at natural phase boundaries, which steer development without interrupting in-progress work. 
Human users can file issues that manager agents consume at subsequent phase boundaries, mimicking the natural workflow of the real-world software development process.

We evaluate \tbc on four real-world software projects over days of continuous operation, along with benchmark-based evaluations to catch team adaptation, verification-driven quality improvement, failure recovery, and practical lessons for future multi-agent systems.

The contributions of this paper are:
\begin{enumerate}
    \item We design and implement \tbc, an open-source orchestration framework for persistent, milestone-driven multi-agent software development with asynchronous oversight.
    \item We introduce a self-organizing team model in which managers hire, specialize, reassign, and retire workers as milestone demands evolve.
    \item We evaluate \tbc on four long-running projects and five hard ProjDevBench problems, and we derive design lessons for continuous multi-agent software development.
\end{enumerate}

\section{Related Work}
\label{sec:related}

\textbf{LLM-based code generation.} LLMs have demonstrated strong capabilities in code generation, progressing from function-level synthesis~\cite{chen2021codex, li2022alphacode, austin2021program} to repository-level tasks~\cite{jimenez2024swebench, zhang2023repocoder, yang2024swe}. Early work (e.g., Codex~\cite{chen2021codex}, AlphaCode~\cite{li2022alphacode}) demonstrated that AI could solve competitive programming problems and generate correct implementations from natural-language descriptions. Generative AI's coding capabilities are being rapidly commercialized through tools such as GitHub Copilot~\cite{github2026copilot}, Claude Code~\cite{anthropic2025claudecode}, and OpenAI Codex~\cite{openai2025codex}. Subsequent models, including StarCoder~\cite{li2023starcoder}, CodeLlama~\cite{roziere2024codellama}, and DeepSeek-Coder~\cite{guo2024deepseek}, improved generation quality through larger training corpora and specialized fine-tuning.
These advances provide the per-invocation generation capability that \textsc{TheBotCompany}'s agents rely on, but they do not address coordination, persistence, or quality assurance across extended development efforts, which motivate a \tbc{}'s orchestration layer.

\noindent\textbf{Single-agent software development systems.}
A growing body of work focuses on single-agent systems that autonomously perform software engineering tasks. SWE-Agent~\cite{yang2024swe} introduced an agent-computer interface that enables an LLM to navigate repositories, edit files, and run tests to resolve GitHub issues, driving rapid progress on SWE-bench Verified~\cite{jimenez2024swebench}. Later work (e.g., OpenHands~\cite{wang2024openhands}, RepairAgent~\cite{bouzenia2025repairagent}) generalized this model with broader agent architectures, sandboxed execution, and iterative program repair. However, all these systems operate within a single invocation session per task. Each run starts from a clean state, addresses one issue, and terminates, with no mechanism to maintain context across tasks, learn from prior resolutions, or coordinate multiple tasks toward a coherent development objective. Quality assurance remains limited to executing pre-existing test suites associated with the target issue, and no system implements an independent verification stage that evaluates whether produced changes satisfy broader project-level goals---a gap that \textsc{TheBotCompany}'s verification phase directly addresses.

\noindent\textbf{Multi-agent software development systems.}
Multi-agent frameworks address the coordination gap by organizing LLM-powered agents into collaborative teams that mirror human software organizations. ChatDev~\cite{qian2024chatdev} introduced a simulated software company with role-specialized agents (CEO, CTO, programmer, tester) communicating through structured dialogue along a waterfall pipeline. Later frameworks refined this model in different ways. For example, MetaGPT~\cite{hong2024metagpt} replaced freeform chat with shared artifacts governed by Standard Operating Procedures, AgileCoder~\cite{nguyen2024agilecoder} introduced sprint-based iterative refinement, and ALMAS~\cite{tawosi2025almas} aligned agents with agile roles to cover end-to-end software lifecycle stages. In all these systems, however, team composition is fixed at design time: roles are predetermined by the framework designer and remain static throughout execution, preventing the system from adapting its workforce as project demands evolve.

\noindent\textbf{Dynamic agent team composition.}
A growing body of work explores dynamic agent composition beyond fixed-role teams. AgentVerse~\cite{chen2024agentverse} and AutoAgents~\cite{chen2023autoagents} adjust group membership or generate custom teams during task-solving based on progress or task descriptions. DyLAN~\cite{liu2024dylan} refines this with a principled agent selection algorithm based on unsupervised importance scores. Self-Organized Agents~\cite{ishibashi2024self} introduces hierarchical spawning, where ``mother'' agents create specialized ``child'' agents scaled to problem complexity. MegaAgent~\cite{wang-etal-2025-megaagent} and Think-on-Process~\cite{lin2024think} go further, generating agent roles and development processes entirely from project requirements without predefined SOPs. ChatDev's Experiential Co-Learning~\cite{qian2023colearning} and Iterative Experience Refinement~\cite{qian2024ier} take a complementary approach, transferring task-solving experience across a series of projects. These systems represent the closest precursors to \textsc{TheBotCompany}'s self-organizing approach. However, all operate within a single execution session or batch of independent tasks: none maintain team state across development cycles, adapt workforce composition in response to verification feedback from prior milestones, or implement an independent verification stage that can reject deliverables and trigger corrective iterations. 

To the best of our knowledge, \tbc is the first framework to combine all three: (1) team persistence \emph{across} milestones rather than within a single session, (2) workforce adaptation driven by verification feedback from prior milestones, and (3) an independent adversarial verification phase that can reject deliverables and trigger corrective cycles. Prior dynamic-composition systems provide at most one of these in isolation, which is the gap our design and evaluation target.

\section{\tbc}
\label{sec:design}

\subsection{Architecture Overview}
\label{sec:architecture}

\begin{figure*}[t]
\centering
\includegraphics[width=0.8\textwidth]{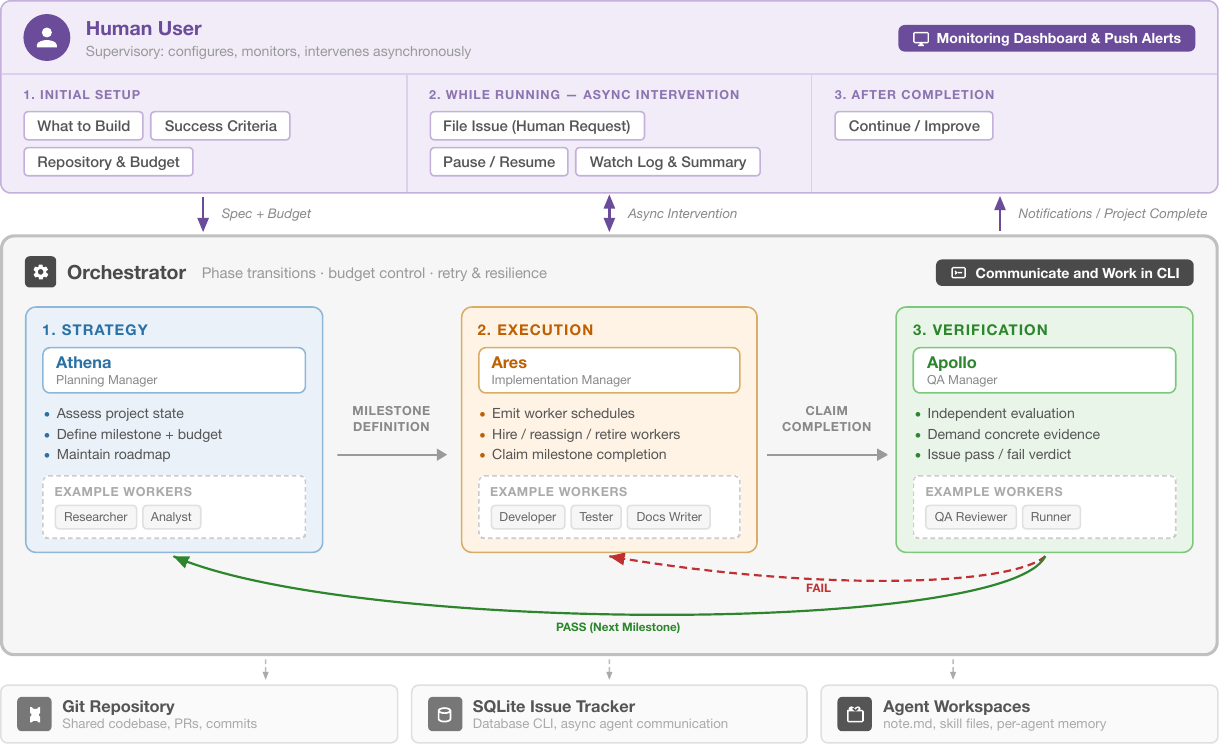}
\caption{Architecture of \textsc{TheBotCompany}. The human monitors the system through a dashboard connected to the orchestrator, which drives a three-phase milestone lifecycle (\strategy $\to$ \execution $\to$ \verification). Manager agents govern each phase and dynamically hire/retire worker agents. Solid arrows show normal progression; dashed arrows show recovery paths. All agents access shared infrastructure through a common bus.}
\label{fig:arch}
\vspace{-5mm}
\end{figure*}

\tbc{} organizes continuous development around these concepts (see \autoref{fig:arch} for an overview):

\textbf{A project} corresponds to a software repository under active development. Each project maintains persistent memory across sessions through a layered storage model (orchestration state, milestone history, an issue tracker with inter-agent comments, cycle reports, and a cost ledger), allowing it to survive restarts, resume mid-milestone, and provide agents with a complete, auditable record of prior decisions. Multiple projects run simultaneously on a single \tbc{} instance, each progressing independently with fully isolated memory.

\textbf{A milestone} is the fundamental unit of work in \tbc, representing a discrete development objective. As real-world software projects are too large for any single agent invocation to address end-to-end, milestones break a long-term project goal into manageable pieces that can each be planned, implemented, and verified independently. A dedicated agent team (\textbf{\athena}) is responsible for milestone planning. Milestones are organized hierarchically using dotted numbering (e.g., 1, 1.1, 1.1.2). The hierarchical design allows the system to automatically adapt milestone granularity to the agents' capabilities by breaking unachievable milestones into smaller tasks.

\textbf{The orchestrator} is the central coordination layer, implemented in JavaScript and running on Node.js. It drives each project through development cycles (see \S\ref{sec:lifecycle}), invoking agents according to the current phase, tracking progress, and managing phase transitions. The orchestrator does not make software engineering decisions itself; rather, it provides the scheduling infrastructure that enables agents to operate autonomously within a structured lifecycle. It is responsible for enforcing time limits, detecting failures, and maintaining persistent state across cycles.

\textbf{The harness} is responsible for directly interacting with LLMs. We use the \texttt{pi-ai} library~\cite{mariozechner2026pi-ai} for unified multi-provider LLM access, streaming, tool calling, and per-call token tracking. On top of \texttt{pi-ai}, \tbc{} adds automatic context compaction, tiered model routing, retry logic with exponential backoff, and structured directive parsing from agent responses.

\textbf{Agents} are the autonomous actors that perform all software engineering work. Each agent is an LLM-based coding agent with access to a project's repository, command-line tools, and a structured communication channel backed by an internal issue tracker. Agents are divided into two categories:

\textbf{Manager agents} are three permanent agents, each assigned to a distinct phase of the milestone lifecycle: \textbf{\athena} (planning), \textbf{\ares} (implementation), and \textbf{\apollo} (verification)\footnote{We use names from Greek mythology for the three manager roles throughout this paper for clarity of exposition.}. Each milestone passes through all three managers in sequence. Managers make strategic and organizational decisions by defining milestones, assembling teams, assigning tasks, and evaluating outcomes. 

\textbf{Worker agents} are the autonomous actors that perform all assigned software engineering work. Each worker is an LLM-based coding agent, and managers may assign different models to workers. A worker is visible only to their manager, enforcing phase-scoped team isolation. Workers perform the actual software engineering tasks: writing and refactoring code, running test suites, opening pull requests, and producing artifacts. 
The orchestrator executes scheduled workers serially within each cycle, as the goal of \tbc is long-horizon, limited-human-involvement software development, where development time is not critical. 

\textbf{Team management} is achieved by managers creating, assigning, and retiring workers in response to milestone needs. Each worker is defined by a \emph{skill file}. A skill file is a Markdown document with YAML front matter that specifies the worker's role (e.g.,  ``backend developer,'' ``test engineer,'' ``documentation writer''), its reporting manager, LLM model tier, and tool permissions. The Markdown body records responsibilities, conventions, and known pitfalls and is updated by the owning agent across cycles. Skill files persist in the project workspace across cycles.

\textbf{Budget management} enforces a user-configured 24-hour rolling spending limit, as autonomous agents invoking commercial LLM APIs can incur substantial costs if left uncontrolled. When a project falls below a safety threshold, the orchestrator introduces adaptive sleep intervals between agents to spread expenditure over the remaining time horizon; once the budget is fully exhausted, it sleeps until the budget replenishes at the start of the next window before resuming automatically.

\textbf{Agent communication} is achieved within \tbc through an issue system (similar to GitHub issues and comments).

\textbf{The human user} configures a project by answering two specification questions: \textit{(1) What do you want to build?} and \textit{(2) How will you determine whether the project is successful?}, along with a repository path and budget. Beyond this setup, users can monitor, supervise, and raise issues in \tbc to provide guidance.

\subsection{Three-Phase Milestone Lifecycle}
\label{sec:lifecycle}

Software development in \tbc is organized around \emph{milestones}, discrete units of work with defined objectives and cycle budgets. Projects progress through a repeating three-phase lifecycle: \textbf{\strategy}, \textbf{\execution}, and \textbf{\verification} (see Algorithm~\ref{alg:orchestrator} for the main loop of \tbc), modeled on the plan\,--\,implement\,--\,review cycle of real-world software engineering practice for improved multi-agent performance~\cite{3766078.3766304}. 

\begin{algorithm}[t]
\footnotesize
\caption{Orchestrator main loop.}
\label{alg:orchestrator}
\begin{algorithmic}[1]
\Require Project $P$
\Loop
    \State $milestone, budget \leftarrow \text{\athena's Team.Plan}(P)$
    \Repeat
        \State $\text{\ares's Team.Execute}(P, milestone, budget, cycle)$
        \State $verdict \leftarrow \text{\apollo's Team.Verify}(P, milestone)$
        \If{$verdict \neq \textsc{Pass}$}
            \State $budget \leftarrow \lfloor budget / 2 \rfloor$
        \EndIf
    \Until{$verdict = \textsc{Pass}$ \textbf{or} $budget = 0$}
\EndLoop
\end{algorithmic}
\end{algorithm}

\textbf{The \strategy phase.}
\athena keeps the project moving toward the user-defined goal. Drawing on the project specification, the issue tracker (which may contain user intervention requests), the codebase, and any feedback from prior verification failures, \athena assesses overall progress, reconciles new requirements with the existing plan, defines a concrete next milestone with a title, description, and \emph{cycle budget}, and updates a persistent roadmap.

\textbf{The \execution phase.}
\ares receives the milestone definition and is responsible for achieving it within the allocated cycle budget. Each cycle, \ares assesses progress and emits a \emph{schedule}, which is an ordered list of workers and their tasks. 
When \ares judges the milestone complete, it issues a \emph{completion claim} and the orchestrator transitions the project to the \verification phase. If the budget expires before completion, control returns to the \strategy phase.
Algorithm~\ref{alg:cycle} details how \ares's team executes a milestone. \athena's and \apollo's teams follow a similar cycle structure but without a cycle budget, as they complete in a single cycle.

\begin{algorithm}[t]
\footnotesize
\caption{\ares's team execution cycles. Other teams follow a similar structure without a cycle budget.}
\label{alg:cycle}
\begin{algorithmic}[1]
\Require Project $P$, milestone $m$, budget $b$
\For{$cycle = 1$ \textbf{to} $b$}
    \State $schedule \leftarrow \text{\ares.Assess}(P, m, b, cycle)$
    \If{\ares claims completion}
        \State \textbf{break}
    \EndIf
    \For{each item $i$ in $schedule$}
        \If{$i$ is a delay}
            \State \textbf{sleep}($i.\text{duration}$)
        \Else
            \State $i.\text{worker.Execute}(P, i.\text{task})$
        \EndIf
    \EndFor
\EndFor
\end{algorithmic}
\end{algorithm}

\textbf{The \verification phase.}
\apollo independently evaluates whether the claimed milestone has been achieved. It may assemble a verification team to run tests, inspect code, or perform quality checks, then issue a \textbf{pass} or \textbf{fail} verdict with structured feedback. On a pass, the system advances to the next \strategy phase. On a fail, if remaining budget permits, the system returns to \execution with the budget halved and \apollo's feedback, giving the execution team a bounded opportunity to address deficiencies. The budget halves on each subsequent failure, progressively tightening the fix window; once exhausted, the milestone is marked failed, and control returns to \athena. 
Halving bounds the correction attempts for a milestone with initial budget $b$ to at most $\lfloor \log_2 b \rfloor$ rounds, giving a deterministic upper bound on per-milestone cost and preventing indefinite cycling on contested verifications. The rule follows the exponential-backoff convention of retry scheduling~\cite{10.1145/360248.360253} rather than an optimality argument; we adopt it as a heuristic with a bounded-cost guarantee, not a claim of convergence to an optimal fix schedule.
\apollo is designed to be \emph{adversarial} to demand concrete evidence, such as test output or CI results, rather than accepting claims at face value, mirroring the separation between development and QA in human software organizations.

\textbf{Structured directives.}
Manager agents communicate decisions to the orchestrator through structured directive blocks embedded in their natural language responses.  These directives follow a defined schema and enable machine-parseable coordination:

\begin{itemize}
    \item \textit{Milestone directives} (\athena) specify the milestone title, description, and cycle budget.
    \item \textit{Schedule directives} (\ares and \apollo) specify workers, their tasks, optional per-agent delays, and visibility modes.
    \item \textit{Completion claims} (\ares) signal that the milestone objectives have been met.
    \item \textit{Verification verdicts} (\apollo) record pass/fail decisions with structured feedback identifying specific deficiencies.
    \item \textit{Project completion directives} (\athena) signal that the project's ultimate goal has been fully achieved, causing the orchestrator to pause and notify the user.
\end{itemize}

The orchestrator parses these directives to drive phase transitions and schedule worker invocations. Manager agents do not communicate with each other directly. Instead, all cross-phase information is mediated by the orchestrator. Specifically, verification feedback from \apollo is injected by the orchestrator into \ares's prompt context at the start of each fix round, and each manager receives the current milestone definition and relevant orchestration state in its context when its phase begins. This design keeps coordination logic within the agents' natural language output while providing the structure needed for automated orchestration.

\subsection{Agent Teaming}

\textbf{Self-organizing agent teams.} Only the three manager roles are permanent, while the worker team is entirely at each manager's discretion. Workers are defined by \emph{skill files}. Each manager may only hire and schedule workers who report to them: \athena may assemble a temporary research team, \ares composes developers and testers, and \apollo staffs QA specialists. We find that the semi-structured design best combines the human wisdom (predefined structure) and AI intelligence (guides concrete worker jobs). The design can maintain a reasonable workflow while keeping the team flexible to the task at hand.

\textbf{Information control.}
Managers control not only \emph{what} workers do but also \emph{what they see}. When scheduling a worker, a manager specifies a \emph{visibility mode}: \textbf{Full} exposes all open issues; \textbf{Focused} exposes only a manager-specified subset (reducing irrelevant context); and \textbf{Blind} provides no issue board context, useful for independent verification tasks where prior discussion might bias the worker. The orchestrator propagates the chosen mode via an environment variable injected into the agent's invocation, which the \texttt{tbc-db} CLI respects when responding to issue queries.

\textbf{Coordination.} Agent coordination is backed by a per-project SQLite database. SQLite is preferred over cloud-hosted trackers (e.g., GitHub Issues) for three reasons: (1) local queries complete in microseconds versus hundreds of milliseconds for remote API calls; (2) cloud rate limits become binding when multiple agents are active across several projects simultaneously; and (3) agent's internal deliberation (e.g., failed attempts, experimental branches) should not pollute the project's public issue tracker or inadvertently trigger notifications by \texttt{@}-mentioning real users who share agent names.

\textbf{Issue tracking.}
Agents communicate asynchronously through an issue tracker modeled on GitHub Issues. The schema stores issues (with title, description, status, and assigned agent), threaded comments, milestone history, complete per-invocation agent reports (including parsed directive blocks), and the active agent registry. Managers open issues to define work items; workers post comments with progress updates; the verification team records findings. Agents access the database through \texttt{tbc-db}, a CLI injected into every agent environment whose commands follow natural conventions (e.g., \texttt{tbc-db issue list -{}-status open}) and produce human-readable output that fits within an agent's context window.

\textbf{Agent-local knowledge base.}
Because each agent invocation starts with a fresh context window, agents face a cold-start problem. To mitigate this, the orchestrator injects a persistent \texttt{note.md} into each agent's context on every invocation. Agents update this file freely to record conventions, known pitfalls, and architectural decisions, and may create additional workspace files (checklists, analysis reports, acceptance criteria logs) that accumulate across cycles into a role-tailored knowledge base.

\subsection{Monitoring Dashboard}
\label{sec:monitor}

\tbc includes a Progressive Web App (PWA) monitoring dashboard served by the orchestrator's built-in HTTP server (no separate backend required), communicating via REST and a Server-Sent Events (SSE) stream for real-time updates. Web Push (VAPID) notifications alert users to milestone completions, verification failures, or system errors on mobile devices even when the browser is closed, supporting unattended operation.

\textbf{Process inspection.} The dashboard displays current phase, active milestone, cycle counter and budget remaining, per-agent status, and cost metrics. A key capability is \emph{live agent streaming}: while an agent runs, its console output, including reasoning, tool calls, file edits, and command execution, is streamed in near real time, letting users understand \emph{how} decisions are reached rather than just their outcomes. Completed reports are archived and filterable per agent; an open-PRs panel aggregates pull requests created by agents across all projects.

\textbf{Intervention and emergency reset.} Users steer development by filing issues in the project tracker, the same mechanism agents use, which manager agents pick up asynchronously in subsequent cycles. Strategic redirection flows through the management hierarchy (\athena consumes user issues first) rather than disrupting in-progress workers. When the system enters a degenerate state (e.g., repeated verification failures cycling without progress), the \emph{bootstrap} operation immediately halts all agents, clears the current milestone, and resets the phase to \strategy so the user can file corrective guidance before \athena re-plans from scratch.

\section{Evaluation}
\label{sec:eval}

We evaluate \tbc through four long-running software deployments over multiple days of continuous, largely unattended operation. 
We treat the four long-running deployments as a \emph{feasibility and behavioral-characterization} study. These studies demonstrate that self-organizing continuous development is achievable, and we use them to characterize how \tbc behaves. The studies cannot claim statistical generality, and we complement them with a controlled benchmark comparison that is externally scored.

\subsection{Research Questions}

\textbf{RQ1: Can a self-organizing multi-agent system make progress toward defined quality objectives over multi-day continuous operation?} 
We measure whether \tbc can make cumulative progress (completing milestones, resolving issues, and growing codebases) without reverting to a degenerate state where cycles produce no useful work.

\noindent\textbf{RQ2: How does the self-organizing multi-agent approach compare against single-agent baselines on hard end-to-end development tasks?}
A core motivation of \tbc's design is that coordinated multi-agent execution should outperform single-agent approaches on development tasks requiring prolonged effort and iterative refinement. We evaluate \tbc against Claude Code on the five hardest problems from ProjDevBench~\cite{lu2026projdevbench}, a recent end-to-end benchmark that evaluates coding agents on system architecture design and functional correctness in repositories with minimal external intervention. The comparison isolates the contribution of multi-agent coordination from model capability by holding the underlying LLM constant in one condition while varying models in the single-agent baseline.

\noindent\textbf{RQ3: How does the verification phase behave in practice, and what defects does it surface?}
The three-phase lifecycle introduces an independent verification step that may reject an ostensibly completed milestone and trigger a fix round. We conduct a systematic log-based analysis of verification behavior across all observed milestones, providing an empirical characterization of whether the verification step catches substantive defects rather than being a passive pass-through.

\noindent\textbf{RQ4: What is the cost profile of continuous multi-agent development, and how does token efficiency compare to single-agent baselines?}
Autonomous development using commercial LLM APIs incurs a direct monetary cost that scales with system complexity. We analyze the cost structure of \tbc across the projects run in our experiments: cost per milestone, token consumption breakdown by agent role (manager overhead versus worker productivity), and the effectiveness of the budget-aware scheduling mechanism in preventing overruns. For comparative purposes, we report total token consumption for ProjDevBench problems across \tbc and Claude Code, providing a quantitative measure of cost-efficiency relative to single-agent execution on identical tasks.

\noindent\textbf{RQ5: What are the dominant failure modes, and how effective are the recovery mechanisms?}
Long-running systems experience diverse failures, including agent timeouts, model errors, incorrect code, and infrastructure disruptions. We categorize observed failures, measure their frequency, and evaluate how effectively the recovery mechanisms (timeout escalation, consecutive-failure detection, cycle accounting, commit-early policy) preserve progress and prevent resource waste.

\subsection{Experimental Setup}

\subsubsection{Subject Projects.}

We evaluate \tbc on two categories of experiments: a standardized benchmark and long-running real-world projects.

\noindent\textbf{Benchmark evaluation.} We select the five problems designated as the hardest in ProjDevBench~\cite{lu2026projdevbench}, an end-to-end coding benchmark that evaluates agents on full project development from requirements, assessing system architecture design, functional correctness, and iterative refinement. Each problem targets a C++ implementation and is evaluated on two dimensions: an execution score (automated Online Judge (OJ) testing, 80 points maximum) and an overall score (100 total) that adds LLM-assisted code review (20 points). We run \tbc under identical conditions to those prescribed by ProjDevBench, where coding agents can retry when an OJ test fails, up to a predefined maximum number of attempts.

\noindent\textbf{Long-running projects.} We choose four projects of varying character, complexity, and implementation language. 

\begin{itemize}
\item \textbf{Project A} (M2Sim): A cycle-accurate, execution-driven, Apple M2 CPU simulator written in Go, targeting correct execution of all benchmarks in the PloyBench~\cite{pouchet2012polybench} and PARSEC~\cite{10.1145/1454115.1454128} suites with timing estimation error below 20\% on average and 50\% per benchmark.

\item \textbf{Project B} (GroundDB): A from-scratch, Python relational database engine with no external database or SQL parsing dependencies, targeting TPC-H benchmark~\cite{dreseler2020quantifying} correctness.

\item \textbf{Project C} (RustLaTeX): A \LaTeX{} compiler in Rust, targeting document parsing, layout computation, and PDF output similarity with other \LaTeX{} compiler outputs.

\item \textbf{Project D} (PyInterpreter): A Python interpreter implemented in C++. Note that Project D is drawn from the ProjDevBench problem set, providing a direct bridge between the benchmark evaluation and the long-running project track.

\end{itemize}

These projects span four implementation languages (Go, Python, Rust, C++) and four domains (systems simulation, database engines, tooling, compiler/interpreter), covering a range of project durations and complexity levels. 

Each project is configured with a brief (20--80 words) natural-language specification answering only \emph{what to build} and \emph{how success is defined}, mirroring realistic usage in which users give short instructions and refine them later through issues. The four specifications fall into three success-criterion regimes: \textbf{executable oracles} authored by the user (Project~A: PloyBench~\cite{pouchet2012polybench}/PARSEC~\cite{10.1145/1454115.1454128} timing bounds; Project~B: TPC-H queries), an \textbf{online-judge oracle} reused from ProjDevBench (Project~D), and a \textbf{qualitative target} (Project~C: \emph{``binary-identical PDF''}).
As an example, the complete M2Sim (Project~A) specification is:
\begin{quote}\small \textbf{What to build.} M2Sim is a cycle-accurate, execution-driven Apple M2 CPU simulator built on the Akita framework~\cite{jannat2026akita}.\\ \textbf{Success criterion.} Support simulating the whole PloyBench benchmark suite and the PARSEC benchmark suite with 20\% average time-estimation error; the maximum error of any single benchmark must not exceed 50\%, where $error=\lvert sim-hw\rvert/min(sim,hw)$. \end{quote}

\subsubsection{Baselines.}
For the ProjDevBench evaluation, we compare \tbc against Claude Code instantiated with five model backends evaluated in the original benchmark, spanning both closed- and open-weight frontier models. \tbc is restricted to Claude Sonnet~4.5 as its backend, prohibiting other models to ensure a fair comparison. All agents invoke their models with provider-default decoding parameters (temperature, top-$p$, etc.); as default decoding is both the fair-comparison condition against Claude Code under the same base model and the deployment-realistic baseline a practitioner would adopt without prior tuning data. ProjDevBench is chosen as the primary evaluation vehicle because it shares the same operational premise as \tbc: given a repository initialized from scratch, the system must develop it autonomously with only limited feedback from an online judge, whereas the majority of existing code agent benchmarks (e.g., SWE-bench~\cite{jimenez2024swebench}) focus on resolving discrete, pre-existing issues in established codebases, a fundamentally different task structure. 

For the long-running project evaluation, no equivalent single-agent baseline exists, as running multi-day development sessions with Claude Code on the same proprietary projects under controlled conditions was not feasible; cost and workflow analysis for these projects is therefore self-contained rather than comparative.

\subsubsection{Intervention Protocol.} 
\label{subsubsec:intervention}
To test whether \tbc sustains continuous operation without human guidance, and to keep runs reproducible, we supplied \emph{no} human instructions during the four long-running projects and the benchmark runs beyond the initial two-question specification. The sole exception is ProjDevBench's OJ feedback, which the benchmark protocol permits. We route this feedback through the same asynchronous human-in-the-loop channel a real user would use: after each completion claim, we write the OJ response into the project issue tracker as a filed issue and resume the run, so \athena consumes it at the next \strategy phase exactly as it would a user-filed issue. This exercises the async oversight channel end-to-end (analyzed in \S\ref{subsubsec:rq2}) while keeping the long-running projects fully unattended.

\subsubsection{Metrics.}

\begin{itemize}
\item \textbf{Progress metrics}: Milestones completed, commits, lines of code added or modified, test cases added, and benchmark scores (execution score and overall score for ProjDevBench problems).
\item \textbf{Verification behavior}: \apollo invocation count, verification rejection rate, defect categories identified at rejection, fix-round iteration count, and milestone-level rework patterns extracted from orchestrator logs.
\item \textbf{Cost and token metrics}: Total token consumption, cost per milestone, token breakdown by agent role (manager versus worker), and budget scheduling effectiveness.
\item \textbf{Reliability metrics}: Timeout rate, consecutive failure events, cycle waste ratio (failed cycles divided by total cycles), and mean cycles to recovery per failure episode.
\end{itemize}

All metrics are collected automatically from the SQLite communication stores, cost tracking logs, Git history, and server logs maintained by \tbc during operation.

\subsection{Results}

All four long-running projects and all five ProjDevBench runs reached an agent-claimed success state within their cycle budgets; the analyses below characterize \emph{how} that state was reached, including the rework and recovery behavior along the way.

\subsubsection{RQ1: Continuous Development Effectiveness.}

Table~\ref{tab:rq1-summary} summarizes the four long-running deployments. Across all projects, \tbc completed 164 milestones and 616 agent cycles spanning 142 wall-clock hours, reaching the primary quality objective in every case without human intervention.
No project exhibited the degenerate behavior we guarded against. Cycle waste ratios were low and only occurred in two projects: 5.3\% in M2Sim, and in RustLaTeX, 27 fix-round milestones account for a 6.9\% waste ratio. The projects also show that \tbc adapts to widely varying task complexity without structural change: GroundDB completed in 17 cycles with a tight correctness oracle, while RustLaTeX required 392 cycles as each layout change interacted with competing rendering constraints. Both trajectories exhibit forward progress rather than oscillation.

\begin{table*}[t]
\footnotesize
\caption{Long-Running Project Deployments.}
\label{tab:rq1-summary}
\setlength{\tabcolsep}{3.5pt}
\begin{tabular}{llrrrrrrr} 
\toprule 
Project & Language & Milestones & Cycles & Cost & Duration (hours) & Cycle/Milestone & \$/Milestone & hour/Milestone \\ 
\midrule 
A (M2Sim) & Go & 12 & 105 & \$527 & 49.1 & 8.8 & 43.9 & 4.09 \\ 
B (GroundDB) & Python & 4 & 17 & \$33 & 2.6 & 4.3 & 8.25 & 0.65 \\ 
C (RustLaTeX) & Rust & 95 & 392 & \$700 & 62.9 & 4.1 & 7.37 & 0.66 \\ 
D (PyInterpreter) & C++ & 53 & 102 & \$261 & 27.8 & 1.9 & 4.92 & 0.52 \\ 
\bottomrule 
\end{tabular}
\end{table*}

\finding{RQ1}{TheBotCompany sustained forward progress across all four long-running projects, spanning four languages (Go, Python, Rust, C++) and three architectural paradigms, completing 164 milestones, 616 cycles, and 142 wall-clock hours without human intervention beyond initial project configuration.}

\subsubsection{RQ2: Comparison with Single-Agent Baselines.}
\label{subsubsec:rq2}

Table~\ref{tab:projdevbench} reports scores on the five hard problems from ProjDevBench. \tbc (sole Claude Sonnet~4.5) achieves a mean execution score of 53.2 and overall score of 70.0, leading all five single-agent baselines. The gap over Claude Code with the \emph{same} underlying model (Sonnet~4.5: 43.6 exec, 59.3 overall) is +9.6 and +10.7 points respectively, directly attributable to multi-agent coordination rather than model capability, since the LLM backend is held constant. 

The advantage is most pronounced on the hardest problems. On P3 and P17, every single-agent system scores near zero ($\leq$7.5), while \tbc scores 48.0 and 22.4---problems where a one-shot session cannot traverse the implementation space that \tbc's milestone structure explores iteratively. On P16, the result reverses: \tbc scores 62.9 execution versus 80.0 for all Claude Code baselines. Post-hoc analysis of P16 logs reveals that its specification admits a compact, self-contained solution, while \tbc's multi-milestone decomposition introduced architectural layering that added verification overhead without proportional benefit, suggesting that multi-agent coordination has diminishing returns when the one-shot solution space is already effective.

\begin{table}[t]
\footnotesize
\caption{ProjDevBench Results: Execution Score (max\,80) and Overall Score
(max\,100). \tbc (BotCo) uses Claude Sonnet~4.5 as its sole backend during these tasks. All other models run in Claude Code. Bold indicates column maximum.}
\label{tab:projdevbench}
\setlength{\tabcolsep}{3.5pt}
\begin{tabular}{lrrrrrr}
\toprule
\# & BotCo & Sonnet~4.5 & DeepSeek-v3.2-exp & GLM-4.6 & GPT-5 & Kimi-K2 \\
\midrule
\multicolumn{7}{l}{\textit{Execution Score (max 80)}} \\
P3   & \textbf{48.0} & 0.0  & 0.0  & 0.0  & 0.0  & 0.0  \\
P4   & 61.4 & \textbf{68.2} & 9.0  & 7.2  & 43.8 & 0.0  \\
P15  & \textbf{71.1} & 62.2 & 44.4 & 62.2 & \textbf{71.1} & 62.2 \\
P16  & 62.9 & \textbf{80.0} & \textbf{80.0} & \textbf{80.0} & \textbf{80.0} & \textbf{80.0} \\
P17  & \textbf{22.4} & 7.5  & 7.5  & 7.5  & 7.5  & 0.0  \\
\midrule
Avg. & \textbf{53.2} & 43.6 & 28.2 & 31.4 & 40.5 & 28.4 \\
\midrule
\multicolumn{7}{l}{\textit{Overall Score (max 100)}} \\
P3   & \textbf{66.7} & 18.0 & 18.0 & 18.0 & 18.7 & 18.7 \\
P4   & 69.8 & \textbf{78.6} & 18.6 & 18.4 & 54.6 & 0.0  \\
P15  & 89.8 & 80.9 & 62.4 & 80.9 & \textbf{90.4} & 80.9 \\
P16  & 82.2 & 98.0 & \textbf{98.7} & 98.0 & 97.3 & 97.3 \\
P17  & \textbf{41.4} & 21.0 & 26.0 & 26.0 & 24.5 & 19.0 \\
\midrule
Avg. & \textbf{70.0} & 59.3 & 44.7 & 48.3 & 57.1 & 43.2 \\
\bottomrule
\end{tabular}
\end{table}

\finding{RQ2}{\tbc (Sonnet~4.5) leads all single-agent baselines on ProjDevBench, gaining +9.6/+10.7 points over the same-model baseline (Sonnet~4.5) with Claude Code on execution/overall score. The advantage concentrates on high-difficulty problems where iterative multi-milestone refinement is essential.}

Because ProjDevBench permits bounded retries with OJ feedback, RQ2 also exercises \tbc's asynchronous oversight channel: each OJ verdict is filed as a project issue and consumed at the next \strategy phase (\S\ref{subsubsec:intervention}). Across all five problems, \athena responded by triggering fresh planning and implementation rather than treating the verdict as a passive signal; the compact P15 redesign (\S\ref{subsubsec:output-scale}) is one outcome of this loop.

\subsubsection{RQ3: Verification Phase Behavior.}
\label{subsubsec:rq3}

\begin{table}[t]
\footnotesize
\caption{Verification Phase Behavior Across Projects (RQ3).
Reject rate\,=\,verify\_fail / total \apollo verdicts.}
\label{tab:rq3-verification}
\setlength{\tabcolsep}{4pt}
\begin{tabular}{lrrl}
\toprule
Project & \apollo Invocation Count & Reject & Dominant Failure Mode \\
\midrule
A (M2Sim)   & 19 &  5.3\% & Accuracy regression \\
B (GroundDB) &  4 &  0.0\% & NA \\
C (RustLaTeX)& 97 & 18.6\% & Similarity regression \\
D (PyInterpreter)  & 27 &  0.0\% & NA  \\
\bottomrule
\end{tabular}
\vspace{-5mm}
\end{table}

Table~\ref{tab:rq3-verification} reveals that \apollo's rejection rate varies systematically with task specification clarity, not with project scale. Projects with exact, automatically evaluable acceptance criteria (Project B and D) produced 0\% rejection: \apollo acted as a confirmatory gate rather than a corrective one (i.e., it confirmed already-passing work rather than blocking and triggering rework cycles). Projects with iterative quality targets (Project C: pixel similarity compared with \texttt{pdflatex}'s~\cite{thanh2004pdftex} output PDF thresholds that shift per milestone; Project A: timing-accuracy bounds across 30 benchmarks) produced substantial rejections, with \apollo catching regressions that would otherwise have shipped (18.6\% of verifications in RustLaTeX and 5.3\% in M2Sim).

Project~C further demonstrates that \tbc can autonomously construct a grounded evaluation target when the user's specification is ill-defined. The project specification stated \emph{``generate binary identical PDF''}; \athena determined early that strict binary identity was unachievable given font-rendering and layout nondeterminism in the toolchain. Rather than stalling, \athena defined a tractable proxy at milestone M29: pixel similarity against \texttt{pdflatex}'s output on the same \texttt{.tex} files. The M29 issue reads: \emph{``Establish the ability to compare our compiler's output against pdflatex output in CI. This is critical for measuring progress toward the `binary identical PDF' project goal.''} This self-defined metric then served as the acceptance criterion for all 97 subsequent verification cycles. The project terminated at 99.21\% average pixel similarity after 95 milestones with an explicit self-aware stopping statement: \emph{``All strategic improvement approaches have been exhausted\ldots{} The remaining 2.12\% gap\ldots{} is dominated by the mega-line layout problem which cannot be fixed without regression---every approach tried (9+ confirmed anti-patterns) consistently regresses similarity.''}

In Project~C, the 18.6\% rejection rate is not evidence of wasted effort: \apollo's rejections directly drove the accumulation of a growing list of documented \emph{anti-patterns} in \athena's planning context---approaches confirmed to cause pixel-similarity regressions. This list constrained \athena's future search space across subsequent milestones. 27 dedicated fix-round milestones produced measurable recovery: whenever a regression milestone was immediately followed by a revert, the similarity score returned to or exceeded the pre-regression baseline.

\noindent\textbf{Case Study (Project C, RustLaTeX).} We trace a five-milestone episode from Project~C that shows the verification-planning feedback loop driving both regression recovery and eventual termination. By M91, eight documented anti-patterns had accumulated in \athena's planning context---layout interventions confirmed to regress pixel similarity---and the baseline stood at 97.88\%.

At M92, \athena reasoned that applying the same layout technique (\texttt{AlignmentMarker} insertion) to \emph{section headings} rather than paragraphs was mechanically distinct from the documented anti-pattern and proposed it as a fix. The milestone passed \apollo's unit-test gate, but \athena's next planning cycle (M93) detected a CI-level similarity regression to 97.51\%. \athena immediately scheduled a revert milestone; \apollo confirmed recovery to the 97.88\% baseline, and the anti-pattern list grew to nine entries. Seventeen cycles later, M95 reproposed the same class of fix with adjusted parameters (27.9\,pt instead of 21\,pt). This time \apollo explicitly failed the milestone:
\begin{quote}
\small\emph{``M95 FAILED: compare.tex regressed from 97.88\% to 97.31\% ($-$0.57\%). AlignmentMarker for section headings is a confirmed dead end regardless of line-height value.''}
\end{quote}
Rather than searching for a tenth variant, \athena declared the project complete, which is a self-aware stopping decision grounded in exhausted hypothesis space.

This episode illustrates three properties of the verification-planning loop:
(1)~\apollo's unit-test gate can miss CI-level regressions, making \athena's planning cycle a necessary second quality layer.
(2)~Anti-pattern knowledge accumulates across milestones and constrains future search, reducing wasted exploration. This is only achievable by sustained teaming and accumulated context. And
(3)~The context-window boundary is a hard limit---\athena misclassified M92's heading-level variant as safe because the prior anti-pattern was documented as ``per-paragraph,'' a distinction that a bounded context cannot reliably track over 90+ milestones.

\finding{RQ3}{The verification phase catches substantive defects. Higher rejection rates correlate with iterative quality targets rather than project scale, and \apollo-driven feedback durably shaped \athena's planning through anti-pattern accumulation.}

\subsubsection{RQ4: Cost Analysis.}

Across all four long-running projects, workers absorbed the dominant share of both cost and time: on average, worker agents consumed 70.6\% of total token cost and 46.5\% of wall-clock time. This distribution is a positive signal for continuous development: high worker concentration indicates that the system allocates the bulk of its budget to actual engineering work rather than coordination overhead. In contrast, \athena's planning share in particular remained modest (9.8\% of cost on average), confirming that milestone orchestration does not crowd out execution capacity as projects scale. 
 
We define \emph{milestone granularity} as wall-clock hours per milestone; it is set by \athena and varies across runs as the planner adapts decomposition to project structure and verification feedback. M2Sim milestones are the coarsest at 4.1\,h each---roughly 
6$\times$ coarser than RustLaTeX (0.66\,h)---which explains its high per-milestone cost (\$43.9) relative to the finer-grained Projects~B, C, and~D (\$4.9--\$8.3 per milestone at 0.5--0.7\,h each).
Spearman correlation between milestone index and per-milestone cost is non-significant in all projects (Project~C: $\rho=-0.06$, $p=0.55$; Project~D: $\rho=0.06$, $p=0.76$), indicating that cost does not systematically escalate as codebases grow.

Table~\ref{tab:rq4-tokens} reports token consumption for \tbc on the five benchmark problems, using output tokens (non-cached model generation) as the primary unit, matching the metric reported for Claude Code baselines in the ProjDevBench benchmark~\cite{lu2026projdevbench}. Because \tbc and Claude Code may have used different numbers of OJ submission attempts per problem, we normalize by per-attempt output-token consumption to enable a fair comparison. Output-token consumption per attempt ranged from 0.53\,M (P15) to 3.50\,M (P17). On a per-attempt basis, \tbc consumed fewer output tokens than Claude Code only on P3 (1.23\,M vs.\ 1.62\,M, $0.76\times$), while consuming more on the remaining four problems---most substantially on P16 (0.81\,M vs.\ 0.15\,M, $5.4\times$) and P17 (3.50\,M vs.\ 1.09\,M, $3.2\times$). Averaged across all problems, \tbc consumed 1.03\,M output tokens per attempt versus 0.62\,M for Claude Code ($1.65\times$ more), reflecting the additional generation cost of coordinating multiple specialized agents within each development cycle. Per-score efficiency varied considerably: P15 achieved 22.3\,K output tokens per execution-score point versus 98.1\,K for P4 ($4.4\times$ worse), indicating that cycle count, rather than problem difficulty alone, is the primary driver of output-token cost. 

\begin{table}[t]
\footnotesize
\caption{Token consumption and wall-clock running duration for \tbc on ProjDevBench problems compared with Claude Code, normalized by number of OJ submission attempts for a fair comparison. Claude Code output-token figures are taken from the ProjDevBench paper~\cite{lu2026projdevbench}.}
\label{tab:rq4-tokens}
\setlength{\tabcolsep}{3pt}
\begin{tabular}{lrrrr}
\toprule
& \multicolumn{2}{c}{\tbc} & \multicolumn{1}{c}{Claude Code} & \\
\cmidrule(lr){2-3}\cmidrule(lr){4-4}
Problem \# & Token/Attempt (M) & Duration & Token/Attempt (M) & Ratio \\
\midrule
P3   & 1.23 & 19.4 hr  & 1.62 & $0.76\times$ \\
P4   & 1.00 & 7.5 days & 0.48 & $2.11\times$ \\
P15  & 0.53 & 23.0 hr  & 0.44 & $1.21\times$ \\
P16  & 0.81 & 9.1 days & 0.15 & $5.41\times$ \\
P17  & 3.50 & 2.1 days & 1.09 & $3.22\times$ \\
\midrule
Average & 1.03 & 0.62 & $1.65\times$ \\
\bottomrule
\end{tabular}
\end{table}

\finding{RQ4}{Worker agents consume 70.6\% of total cost across projects. Per-milestone cost is stable across the project lifetime.  On a per-OJ-submission basis, \tbc consumes on average $1.65\times$ more output tokens than Claude Code (1.03\,M vs.\ 0.62\,M per attempt), reflecting the additional generation cost of multi-agent coordination within each development cycle.}

\subsubsection{Output Scale.}
\label{subsubsec:output-scale}
Table~\ref{tab:rq4-loc} compares the lines of code (LOC) and file count of \tbc's final submissions against the average human submission scale reported in ProjDevBench~\cite{lu2026projdevbench} for the same five problems. \tbc's submissions are broadly proportionate to the expected problem scale: on P3 and P4, generated LOC slightly exceeds the human average (1.48$\times$ and 1.30$\times$ respectively), while P17, the most architecturally complex problem (4,204 human-average lines across 20 files), is matched to within 7\% (3,910 lines). 
The notable exception is P15, where \tbc produces only 377 lines against a 2{,}094-line human average (0.18$\times$). It is a deliberately compact, problem-specific design by \tbc: after its previous submission received an OJ timeout, \tbc adopted a hash-bucket implementation that exploits the P15 specification, optimizing for judged performance rather than for building a complete system (the apparent focus of human submissions); the competitive execution score confirms the solution is strong rather than incomplete.
Across all five problems, \tbc consistently adopts a flatter file structure than human submissions, using between one-third and one-quarter as many files on multi-file problems (P4: 4 vs.\ 18; P17: 6 vs.\ 20), reflecting its tendency to consolidate related logic rather than decompose into fine-grained modules.

\begin{table}[t]
\footnotesize
\caption{Final submission scale for \tbc compared with average human
submission scale from ProjDevBench~\cite{lu2026projdevbench}.
Human figures are averages over human submissions per problem.}
\vspace{-2mm}
\label{tab:rq4-loc}
\setlength{\tabcolsep}{4pt}
\begin{tabular}{lrrrrr}
\toprule
& \multicolumn{2}{c}{\tbc} & \multicolumn{2}{c}{Human Average} & \\
\cmidrule(lr){2-3}\cmidrule(lr){4-5}
Prob. & LOC & Files & LOC & Files & LOC Ratio \\
\midrule
P3  &   705 &  1 &   475.3 &  1 & $1.48\times$ \\
P4  & 2,161 &  4 & 1,659.0 & 18 & $1.30\times$ \\
P15 &   377 &  3 & 2,093.9 &  3 & $0.18\times$ \\
P16 & 1,791 &  7 & 3,139.4 & 11 & $0.57\times$ \\
P17 & 3,910 &  6 & 4,204.3 & 20 & $0.93\times$ \\
\bottomrule
\end{tabular}
\end{table}

\subsubsection{RQ5: Failure Modes and Recovery.}

We derived the failure taxonomy by qualitative thematic analysis~\cite{Braun01012006} of four artifact types---\apollo verification verdicts, issue-tracker records, CI/test output, and milestone reports---coding each distinct failure episode into one category. Two authors coded independently and reconciled disagreements by discussion; the final rubric can be found in~\autoref{tab:rq5-failures}.

\begin{table*}[t]
\footnotesize
\caption{Failure mode taxonomy, rubric, and counts across all experiments (RQ5). Counts are distinct failure episodes.}
\label{tab:rq5-failures}
\setlength{\tabcolsep}{5pt}
\begin{tabular}{@{}l p{6.8cm} rrrr@{}}
\toprule
Failure Mode & Definition & M2Sim & GroundDB & RustLaTeX & PyInterpreter \\
\midrule
Accuracy/similarity regression & Quantitative quality metric (timing error, pixel similarity) drops below its prior baseline & 3 & 0 & 18 & 0 \\
Architectural issue & Design decision requires structural rework rather than a local fix & 1 & 0 & 0 & 0 \\
Missing requirement & Specified capability found unimplemented at verification & 0 & 0 & 5 & 0 \\
Agent execution timeout & Agent invocation exceeds its wall-clock limit & 0 & 0 & 0 & 2 \\
Inter-milestone regression & Completed milestone breaks a previously passing one & 0 & 0 & 0 & 1 \\
Tooling / environment & Infrastructure failure rather than a fault in agent output & 0 & 1 & 0 & 0 \\
\midrule
Total episodes & & 4 & 1 & 23 & 3 \\
\bottomrule
\end{tabular}
\end{table*}

Table~\ref{tab:rq5-failures} shows that failure modes are predominantly domain-specific rather than orchestration failures. The largest failure categories were accuracy/similarity regression (Projects~A and~C), not from deficiencies in agent coordination. Generic software engineering failures (missing requirements, architectural issues) are comparatively rare.

Recovery was effective for all in-system failures. Both \athena execution timeouts in Project~D recovered cleanly in the next cycle: the orchestrator restarted \athena with full project context from the SQLite communication store, and the subsequent invocation defined a valid milestone without re-encountering the timeout. The single tooling failure in Project~B (\texttt{tbc-db} CLI not found at Cycle~1) was self-resolved by \athena switching to direct \texttt{sqlite3} calls, with zero cycle waste.

One failure case represents deliberate non-recovery decisions (see \S\ref{subsubsec:rq3}). In Project~C, a 2.12\% irreducible gap in file similarity was acknowledged by \athena: nine distinct approaches had each been tried and documented as anti-patterns, and no unexplored strategy remained.

The cycle waste ratio (failed cycles / total) was near zero for Projects~A, B, and~D. In Project~C, 27 fix-round milestones represent acknowledged rework cycles; if these are classified as waste, the waste ratio is 27/392\,=\,6.9\%---a modest overhead given the iterative nature of pixel-fidelity optimization. In ProjDevBench problem~P4, approximately 120 late cycles targeted an unresolvable bug, yielding an estimated waste ratio of 31\%.

\finding{RQ5}{Failure modes are domain-specific; orchestration-level failures (timeouts, tooling) are rare and recover within one cycle. Self-aware termination—explicit documentation of irreducible limitations—prevents runaway cost in both the long-running settings.}

\subsection{Threats to Validity}

\noindent\textbf{Internal Validity.} LLM \emph{non-determinism} (non-zero temperature) means rerunning any project would yield different decompositions, artifacts, and quality outcomes; we report single-run results, precluding variance estimation. Researcher authorship of initial project briefs and acceptance criteria also influences development trajectory, though we mitigate this by reporting objective outcomes only (benchmark correctness, test pass rates) rather than handcrafted milestones.

\noindent\textbf{External Validity.} Four long-running projects skewed toward system software (simulator, database engine, compiler, interpreter) with deterministic correctness oracles may not represent the broader distribution of real-world tasks; projects with ambiguous requirements are not evaluated. RQ2 results are partially insulated from selection bias by external OJ scoring comparison with ProjDevBench. All experiments use Claude-family models; whether multi-agent coordination benefits generalize to other LLM backends remains unknown.
 
\noindent\textbf{Construct Validity.} LOC and test count are imperfect quality proxies, supplemented by domain-specific metrics (PloyBench accuracy, TPC-H query coverage, pixel similarity, OJ scores) to reduce reliance on any single measure. \apollo's LLM-based milestone assessment may produce incorrect accept/reject decisions; we partially address this by anchoring acceptance criteria to executable tests over prose evaluation. Cost comparisons between \tbc and single-agent baselines in the long-running track are unavailable, as multi-day Claude Code sessions under identical budgets were not feasible. A related concern is whether the projects merely reproduce existing reference implementations. Two of the four have no public reference to draw on: to our knowledge, no cycle-accurate Apple~M2 simulator (Project~A) and no Rust \LaTeX{} compiler at the target fidelity (Project~C) has been released when we conducted our experiments. Inspecting the orchestrator logs for all four projects, we observe agents building incrementally from an empty repository---defining data structures, implementing components, and adding tests milestone by milestone---rather than transcribing a known solution; web searches inform local decisions (e.g., API usage) rather than supplying a wholesale implementation. Where executable oracles do exist (Projects~A, B, D), they constrain acceptance but not the implementation path.

\section{Discussion}
\label{sec:discussion}

\textbf{Sequential Phase Separation vs. Parallel Competition.}
Although we compared with Claude Code's single-agent mode, the agent teams mode~\cite{anthropic2025agentteams}, which coordinates multiple Claude Code instances under a single team lead, represents an emerging alternative paradigm. 
We did not include a direct comparison because the feature remains experimental and is not publicly available through standard usage at the time we conducted our experiments, and results would be difficult to reproduce and unfair to both systems. Conceptually, however, the architectures differ in a fundamental way. Claude Code's teams favor \emph{parallel, competitive} coordination---agents attempt the same task and the lead selects the best result---whereas \tbc enforces \emph{sequential phase separation}: planning, implementation, and verification are isolated so that each stage cannot be contaminated by the concerns of another. Independent verification is only meaningful when the verifier has no stake in the implementation, a principle that parallels the self-critique loops shown to improve agent quality in Reflexion~\cite{shinn2023reflexion}. We hypothesize that this separation explains the performance pattern observed in RQ2: on compact problems amenable to a single session, single execution is sufficient; on high-difficulty problems requiring iterative refinement, sequential phase separation with persistent verification feedback provides a structural advantage. As such, rigorous empirical validation of this hypothesis, ideally on a shared benchmark with both systems, is an important direction for future work. 
Extending \tbc to run multiple workers per cycle, with the attendant merge and conflict-resolution machinery, could raise a distinct set of research questions (concurrent-edit reconciliation, verification under nondeterministic interleavings, etc.) that we leave to future work.

\textbf{When does multi-agent coordination justify its cost?} 
Our results suggest a practical crossover point. On ProjDevBench problems where the solution space is large and
requires iterative exploration (P3, P17), \tbc outperforms single-agent baselines despite consuming 1.65$\times$ more output tokens per attempt. On problems admitting compact, single-session solutions (P16), the multi-agent overhead reduces both score and cost-efficiency. For practitioners, this implies that multi-agent orchestration is best suited to projects where (1) the implementation cannot be completed in a single session, (2) quality criteria are continuous or iterative rather than binary, and (3) the cost of undetected regressions justifies an independent verification phase. When these conditions do not hold, a single-agent tool is likely more efficient.

\textbf{Human-in-the-loop Oversight Granularity.}
\tbc was designed for largely unattended operation, with human steering channeled through asynchronous issue filing. 
In practice, adding live agent output streaming to the dashboard created an implicit expectation of finer-grained control. For example, users who could watch a worker diverge naturally wanted to correct it mid-cycle. This reflects a tension in mixed-initiative systems~\cite{10.1145/302979.303030}: exposing system internals raises awareness of errors but shifts the perceived locus of control, increasing felt demand for intervention. The appropriate granularity of oversight depends on user expertise and task stakes~\cite{8804457}---a domain expert debugging a numerical simulator has actionable knowledge that a generalist user does not. 
Our experience suggests a tiered model: coarse-grained issue filing for users who prefer autonomy; mid-grained pause/bootstrap for correcting runaway cycles; and potential fine-grained interruption for experts who can productively redirect an in-progress agent. However, how to expose these tiers without overburdening the agent communication infrastructure deserves more future work.
Our evaluation exercised the asynchronous channel for corrective feedback (OJ verdicts filed as issues) but deliberately held requirements fixed for reproducibility. We leave how \tbc interacts and re-plans with mid-project requirement steering from new issues as future work.

\section{Conclusion}
\label{sec:conclusion}

We presented \tbc, an open-source framework for continuous multi-agent software development built around three innovations: a repeating \strategy~$\to$~\execution~$\to$~\verification lifecycle, self-organizing worker teams that managers compose per milestone, and budget-aware scheduling for unattended operation. Evaluation across four real-world projects and five problems yields three findings. First, the verification phase adds corrective value when acceptance criteria are continuous or iterative, but acts as confirmatory overhead when deterministic oracles are available, suggesting that verification cost should be calibrated to specification clarity. Second, self-organizing teams concentrate cost on implementation work, with orchestration overhead remaining stable as projects grow. Third, in a single-run comparison, \tbc achieved higher scores than the same-model single-agent baselines on high-difficulty problems requiring iterative refinement, while single-agent execution was more efficient on problems amenable to single-session solutions. Collectively, our results suggest that structural choices---phase separation, persistent state, and decoupled verification---are important complements to model capability for long-horizon software development.

\section*{Data Availability Statement}
\label{sec:data-availability}
\textsc{TheBotCompany} is an open-source project licensed under MIT. The source code, the four long-running project specifications, and the experiment logs are archived at \url{https://doi.org/10.5281/zenodo.21775778}~\cite{lyu2026thebotcompany}.

\begin{acks}
We thank our anonymous reviewers for their reviews. This work is supported by the National Science Foundation under award no. NSF-2418582 and CCF-2402804, Commonwealth Cyber Initiative under award no. HC-2Q26-037, and William \& Mary's Global Research Institute.
\end{acks}

\bibliographystyle{ACM-Reference-Format}
\bibliography{references}

\end{document}